\documentclass[a4paper,12pt]{scrartcl}
\usepackage{amsmath}
\usepackage{algorithm,algorithmic}
\usepackage{graphicx}
\usepackage{float}
\usepackage{color,dsfont}
\usepackage{clrscode}
\usepackage{hyperref}
\usepackage{blkarray}
\usepackage{pst-node}
\usepackage{mathtools}
\usepackage{authblk}
\usepackage[numbers,sort&compress]{natbib}

\title{Exact numerical calculation of fixation probability and time on graphs}

\date{\normalsize\today}
\begin{document}
\author{Laura Hindersin$^1$, Marius M\"{o}ller$^{1,2}$, Arne Traulsen$^1$, Benedikt Bauer$^{1,3}$}
\affil{\small{$^1$Department of Evolutionary Theory, Max Planck Institute for Evolutionary Biology, D-24306 Pl\"{o}n, Germany} \\
\small{$^2$Institute of Mathematics, University of L\"{u}beck, 23562 L\"{u}beck, Germany} \\
\small{$^3$Evotec AG, D-22419 Hamburg, Germany}}

\maketitle

\noindent

\copyright 2016. This manuscript version is made available under the CC-BY-NC-ND 4.0 license \href{http://creativecommons.org/licenses/by-nc-nd/4.0/}{http://creativecommons.org/licenses/by-nc-nd/4.0/}.

\section*{Abstract}
The Moran process on graphs is a popular model to study the dynamics of evolution in a spatially structured population.
Exact analytical solutions for the fixation probability and time of a new mutant have been found for only a few classes of graphs so far.
Simulations are time-expensive and many realizations are necessary, as the variance of the fixation times is high.
We present an algorithm that numerically computes these quantities for arbitrary small graphs by an approach based on the transition matrix.
The advantage over simulations is that the calculation has to be executed only once. 
Building the transition matrix is automated by our algorithm.
This enables a fast and interactive study of different graph structures and their effect on fixation probability and time.
We provide a fast implementation in C with this note \cite{hindersin:URL:2016}.
Our code is very flexible, as it can handle two different update mechanisms (Birth-death or death-Birth), as well as arbitrary directed or undirected graphs.

\section{Introduction}

The Moran process describes the spread of a newly introduced mutant into a resident population.
Its framework has been extended from a well-mixed population, which can be formulated as a one-dimensional Markov chain, to a process on graphs~\cite{lieberman:Nature:2005,broom:PRSA:2008,adlam:SciRep:2014}, where the configuration of the mutants on the graph has to be taken into account.
The graph represents the spatial structure of the population.
Individuals are located on the vertices and connected via the edges of the graph.
We study the discrete-time Moran process where there is one birth event and one death event per time step.
Mutants have a relative fitness $r>0$ and residents have fitness $1$.
In each time step, individuals have to be picked consecutively for birth and death.
There are several possibilities to implement the details of these updates, however this is not essential for solving the system and is discussed elsewhere~\cite{antal:PRL:2006,zukewich:PlosOne:2013,kaveh:JRSOS:2015}.
We consider only updating rules where birth happens according to fitness and death happens at random.
In Birth-death updating, an individual is selected with probability proportional to its relative fitness.
It produces an identical offspring which replaces a randomly chosen neighbor.
In death-Birth updating, a randomly picked individual dies.
Its neighbors compete with probability proportional to their fitness to fill the empty vertex with their identical offspring.

As we assume no mutations, the process eventually gets absorbed in a homogeneous state with either only wild-type or only mutant individuals.

We use a numerical approach based on the transition matrix to calculate fixation probabilities and times~\cite{grinstead:book:1997,hindersin:PlosCB:2015}. 
First, we explain how to compute the transition probabilities, given the adjacency matrix of the graph.
We make use of the full state space approach introduced in~\cite{voorhees:PRSA:2013b} for Birth-death updating.
Then, we show how to compute the fixation probability and time by solving a linear system.
Finally, we discuss computational limitations and performance.

\section{Software description}
\subsection{Computing the transition matrix from the adjacency matrix}
\subsubsection{Birth-death updating}
Given a graph $G = (V,E)$ with vertices $V=\{1,2,\cdots,N\}$ and edges $E$. An edge $e_{i,j}$ connecting vertices $i$ and $j$ is denoted by $e_{i,j} \in E$.
An entry of the adjacency matrix $A$ is given by
\begin{equation}
A_{i,j} = 
\begin{cases}
1 \, , 	& e_{i,j}\in E \\
0 \, , 	& e_{i,j}\notin E  \ .
\end{cases}
\end{equation}

The graph can be undirected, i.e. $A_{i,j} = A_{j,i}$ for all $i,j \in V$, or directed.
We assume the graph to be (strongly) connected, meaning that there is a (directed) path from vertex $i$ to $j$, for all $i,j \in V$, and simple, i.e. there are no self-loops and parallel edges.
Let $S = \{i_1,i_2,\ldots,i_l\} \in \mathcal{S} = 2^V$ be a state of the Moran process on graph $G$, where $l \in \{1,2,\ldots,N-1\}$ mutants are located at the vertices in $S$, i.e.~we exclude the cases with no mutants or only mutants here.
Now the transition probability from state $S$ to $S \cup \{i_{l+1}\}$, whereby increasing the number of mutants to $l+1$ is given by
\begin{equation}
T_{\{i_1,i_2,\ldots,i_l\} \rightarrow \{i_1,i_2,\ldots,i_l,i_{l+1}\}} = \frac{r}{r l + N-l} \left( \sum_{k=1}^l{\frac{A_{i_k,i_{l+1}}}{\deg_{out}{(i_k)}}} \right) \ ,
\label{eq:BdOneMore}
\end{equation}
where $\deg_{out}{(i_k)}$ is the number of neighbors (successors in the case of directed graphs) of vertex $i_k$, given by the sum of the $i_k$-th row of the adjacency matrix $A$.

Decreasing the number of mutants from state $S$ to state $S\setminus \{i_j\}$, where $j\in \{1,\ldots,l\}$, is given by the following transition probability
\begin{equation}
T_{\{i_1,i_2,\ldots,i_l\} \rightarrow \{i_1,i_2,\ldots,i_{j-1},i_{j+1},\ldots,i_l\}} = \frac{1}{r l + N-l} \left( \sum_{\substack{k=1 \\ i_k\notin S}}^N{\frac{A_{i_k,i_j}}{\deg_{in}{(i_k)}}} \right) \ .
\label{eq:BdOneLess}
\end{equation}

Additionally, the probability to stay in state $S$ is given by $1$ minus the row sum of the transition matrix
\begin{align}
T_{\{i_1,i_2,\ldots,i_l\} \rightarrow \{i_1,i_2,\ldots,i_l\}} = 1 & - \sum_{\substack{l+1=1 \\ i_{l+1}\notin S}}^N{T_{\{i_1,i_2,\ldots,i_l\} \rightarrow \{i_1,i_2,\ldots,i_l,i_{l+1}\}}} \nonumber \\
& - \sum_{j=1}^l{T_{\{i_1,i_2,\ldots,i_l\} \rightarrow \{i_1,i_2,\ldots,i_{j-1},i_{j+1},\ldots,i_l\}}} \ .
\end{align}
The transition probabilities from the states with no mutants or only mutants are $T_{\emptyset \rightarrow \emptyset} = T_{V \rightarrow V} = 1$.
This transition matrix forms the basis to compute the fixation probability and time in Sections \ref{sec:fixprob} and \ref{sec:fixtime}.

\subsubsection{death-Birth updating}
In the Moran process with death-Birth updating, a random individual is selected for death and its neighbors compete for the empty vertex with a probability proportional to their fitness.

Let $m_j$ denote the number of mutant neighbors (predecessors in the case of directed graphs) of a given vertex $i_j$, $j \in \{1,2,\ldots, N\}$. 
Then $m_j = \sum_{k=1}^N{A_{i_k,i_j} \mathds{1}_{\{i_k \in S\}}}$, where $\mathds{1}_{\{i_k \in S\}}$ is the indicator function, being one if $i_k \in S$ and zero otherwise. 
The number of wild-type neighbors (predecessors) of the individual located at vertex $i_j$ are therefore given by $\deg_{in}{(i_j)} - m_{j}$.

Then the transition probabilities are given by
\begin{align}
T_{\{i_1,i_2,\ldots,i_l\} \rightarrow \{i_1,i_2,\ldots,i_l,i_{l+1}\}} &= \frac{1}{N} \frac{r \cdot m_{l+1}}{r \cdot m_{l+1} + \deg_{in}{(i_{l+1})} - m_{l+1}} \\
T_{\{i_1,i_2,\ldots,i_l\} \rightarrow \{i_1,i_2,\ldots,i_{j-1},i_{j+1},\ldots,i_l\}} &= \frac{1}{N} \frac{\deg_{in}{(i_j)} - m_j}{r \cdot m_j + \deg_{in}{(i_j)} - m_j}\\
T_{\{i_1,i_2,\ldots,i_l\} \rightarrow \{i_1,i_2,\ldots,i_l\}} = 1 & - \sum_{\substack{l+1=1 \\ i_{l+1}\notin S}}^N{T_{\{i_1,i_2,\ldots,i_l\} \rightarrow \{i_1,i_2,\ldots,i_l,i_{l+1}\}}} \nonumber \\
& - \sum_{j=1}^l{T_{\{i_1,i_2,\ldots,i_l\} \rightarrow \{i_1,i_2,\ldots,i_{j-1},i_{j+1},\ldots,i_l\}}}
\end{align}

With $T_{\emptyset \rightarrow \emptyset} = T_{V \rightarrow V} = 1$, this transition matrix is used to compute the fixation probability and time for the Moran process with death-Birth updating.

\subsection{Fixation probability}\label{sec:fixprob}

Given the transition matrix from above, the Master Equation for the transition probabilities is given by

\begin{align}
    \phi_{S} = \sum_{R\in \mathcal{S}} T_{S \rightarrow R} \cdot \phi_{R} \quad \quad \forall S \in \mathcal{S}.
    \label{eq:MasterGraph}
\end{align}

For well-mixed populations, which can be represented by a complete graph, the Master Equation has a closed form solution given by a sum over a product of the ratio of transition probabilities~\cite{karlin:book:1975,traulsen:bookchapter:2009}.

Ordering the states by the number of mutants first and by their position next, the transition matrix is given by

\begin{align}
    \mathbf{T}_{2^N \times 2^N} = \left( T_{S \rightarrow R} \right)_{\left( S,R \right)},
    \label{eq:TransMatrix}
\end{align}
where $S$ denotes the starting state and $R$ the target state of the population.

Rewriting this as a block matrix where $\mathbf{Q}_{(2^N-2) \times (2^N-2)}$ is the transition matrix between transient states yields

\begin{equation}
\mathbf{T} = 
\left(
\begin{array}{c|ccc|c}
1 & &\vec{0}^T & & 0 \\
\hline 
 & & & & \\
\vec{p_1} & & \mathbf{Q}& & \vec{p_2} \\
 & & & & \\
\hline
0 & & \vec{0}^T& & 1
\end{array}
\right)
\label{eq:blockMatrix}
.
\end{equation}

Let us write the Master Equation as an Eigenvector problem:

\begin{equation}
\mathbf{T} \vec{\phi} = \vec{\phi} \ ,
\label{eqn:eigen}
\end{equation}
where $\vec{\phi} = (\phi_{\emptyset},\phi_{\{1\}},\ldots,\phi_{\{N\}}, \phi_{\{1,2\}},\ldots,\phi_{\{N-1,N\}},\phi_{\{1,2,3\}},\ldots, \phi_V)^T$.

If we use the block formulation of Eq.~\eqref{eq:blockMatrix}, then Eq.~\eqref{eqn:eigen} is equivalent to

\begin{align}
1  \phi_{\emptyset} + \vec{0}^T  \tilde{\phi} + 0  \phi_V &= \phi_{\emptyset} =0, \label{eq:eigenInnerPart1} \\
\vec{p_1} \phi_{\emptyset} + \mathbf{Q} \tilde{\phi} + \vec{p_2} \phi_V &= \tilde{\phi} , \label{eq:eigenInnerPart2} \\
0  \phi_{\emptyset} + \vec{0}^T  \tilde{\phi} + 1  \phi_V &= \phi_V = 1 \ , \label{eq:eigenInnerPart3} 
\end{align}
where $\tilde{\phi} =   (\phi_{\{1\}},\ldots,\phi_{\{N\}}, \phi_{\{1,2\}},\ldots,\phi_{\{N-1,N\}},\phi_{\{1,2,3\}},\ldots, \phi_{\{2,3,\ldots,N\}})^T$. 
Eqs.~\eqref{eq:eigenInnerPart1} and \eqref{eq:eigenInnerPart3} are always fulfilled. 
Thus, we need only solve Eq.~\eqref{eq:eigenInnerPart2}.
We can subtract $\tilde{\phi}$ on both sides of Eq.~\eqref{eq:eigenInnerPart2} and bring $\vec{p_2}$ to the right side:

\begin{equation}
\label{eq:NumericalApproach}
(\mathbf{Q} - \mathbf{I})  \tilde{\phi} = - \vec{p_2} \ ,
\end{equation}
where $\mathbf{I}$ is the identity matrix.
Now this matrix system can be solved for $\tilde{\phi}$.

\subsection{Fixation time}
\label{sec:fixtime}

\subsubsection{Unconditional fixation time}

Next, we compute the unconditional fixation time by modifying the approach above.
The unconditional fixation time for a state $S\in\mathcal{S}$ is the time it takes starting from that state until the absorbing all-mutant state $S=\{1,2,\ldots,N\}$ or the absorbing no mutant state $S=\emptyset$ is reached.
The unconditional fixation time is recursively given by
\begin{align}
	\tau_S = 1 + \sum_{R \in \mathcal{S}}{T_{S \rightarrow R} \cdot \tau_{R}}  \quad \quad \forall S \in \mathcal{S}.
\label{eq:uncond}
\end{align}
Here, the boundary conditions are $\tau_\emptyset = \tau_V = 0$.
The transition matrix is the same as for computing the fixation probability (cf. Eq.~\eqref{eq:TransMatrix}).
To account for the addition of the one in Eq.~\eqref{eq:uncond}, we add a column of ones and a row of zeros, where the new $(1,1)$ entry is $1$.
\begin{equation}
\mathbf{\mathcal{T}} = \left(\begin{array}{c|c}
    1 & \vec{0}^T \\
    \hline
    \vec{1} & \mathbf{T}
\end{array}\right) = 
\left(
\begin{array}{cc|ccc|c}
1 & 0 & &\vec{0}^T & & 0 \\
1 & 1 & &\vec{0}^T & & 0 \\
\hline 
 & & & & \\
\vec{1} & \vec{p_1} & & \mathbf{Q}& & \vec{p_2} \\
 & & & & \\
\hline
1 & 0 & & \vec{0}^T& & 1
\end{array}
\right)
\label{eq:blockMatrixTime}
.
\end{equation}

We can now write the computation of the unconditional fixation times as formulated in Eq.~\eqref{eq:uncond} as an Eigenvector problem
\begin{align}
    \mathcal{T}\vec{\tau} = \vec{\tau},
\end{align}
where $\vec{\tau} = \left( 1, \tau_{\{\emptyset\}}, \tau_{\{1\}}, \ldots, \tau_{\{N\}}, \tau_{\{1,2\}}, \ldots, \tau_V \right)^T$.

Writing Eq.~\eqref{eq:uncond} in notation of the block matrices as in \eqref{eq:blockMatrixTime}, we obtain
\begin{align}
 1 &= 1 \nonumber\\
 1 + 1  \tau_{\emptyset} + \vec{0}^T  \tilde{\tau} + 0  \tau_V &= \tau_{\emptyset} =0, \label{eq:eigenInnerPartTime1} \\
 1 + \vec{p_1} \tau_{\emptyset} + \mathbf{Q} \tilde{\tau} + \vec{p_2} \tau_V &= \tilde{\tau} , \label{eq:eigenInnerPartTime2} \\
 1 + 0  \tau_{\emptyset} + \vec{0}^T  \tilde{\tau} + 1  \tau_V &= \tau_V = 0 \ , \label{eq:eigenInnerPartTime3} 
\end{align}
where $\tilde{\tau} = \left( \tau_{\{1\}}, \ldots, \tau_{\{N\}}, \tau_{\{1,2\}}, \ldots, \tau_{\{2,3,\ldots, N\}} \right)^T$.
Note, that \eqref{eq:eigenInnerPartTime1} and \eqref{eq:eigenInnerPartTime3} are always true.
Hence, we only need to solve \eqref{eq:eigenInnerPartTime2}.
We subtract $\tilde{\tau}$ and 1 on both side and thus obtain a linear system of equations
\begin{align}
    (\mathbf{Q} - \mathbf{I})  \tilde{\tau} = - \vec{1} \ .
\end{align}
We see that we can use the exact same matrix $\mathbf{Q}$ for the computation of the unconditional fixation times.
Only the vector we solve for changes ($-\vec{1}$ instead of $-\vec{p}_2$).

\subsubsection{Conditional fixation time}

In this section, we explain how to obtain the transition probabilities conditioned on fixation of the mutant to calculate the conditional fixation time.
The recursive equation for the conditional fixation time is similar to Eq.~\eqref{eq:uncond}

\begin{align}
    \tau_{S | fix} = 1 + \sum_{R \in \mathcal{S}}{T_{S \rightarrow R | fix} \cdot \tau_{R | fix}}  \quad \quad \forall S \in \mathcal{S}.
\label{eq:KolmCond}
\end{align}
To compute the conditional transition probabilities, we make use of Bayes' theorem

\begin{align}
    P(A|B) = \frac{P(B|A)P(A)}{P(B)}
\end{align}
and obtain

\begin{align}
    T_{S\rightarrow R | fix} = \frac{p_{fix | R}}{p_{fix | S}} T_{S\rightarrow R} \quad \quad \forall S,R \in \mathcal{S} \ ,
\end{align}
where $p_{fix | S} = \phi_S$ is the fixation probability starting from state $S \in \mathcal{S}$.
We see that for the calculation of the conditional transition probabilities, we need to weight the unconditional transition probabilities by the ratio of the fixation probabilities between the new ($S$) and old ($R$) state~\cite{ewens:TPB:1973,altrock:JTB:2012}.
These fixation probabilities need to be calculated first, making use of the results of Section \ref{sec:fixprob}.
Otherwise, the approach is the same as for the unconditional times.

\subsection{Computational limitations and performance}

\subsubsection{Computational limitations}
The size of the matrix $\mathbf{Q}$ is $(2^N-2) \times (2^N-2)$, where $N$ denotes the population size.
In a naive implementation, the matrix would therefore need $\left(2^N-2\right)^2 \times 8 $ bytes of memory.
For a population of 10 individuals, this would mean 8,355,872 bytes $\approx$ 8.4 MB.
This is not a problem for standard computers nowadays.
But because the matrix grows quadratic exponentially in population size (with $2^{2N}$), a population with 20 individuals would need 8,796,059,467,808 bytes $\approx$ 8796 GB.
This is already way beyond possible today's regular computers.

Most of the entries in the matrix are zero, however \cite{broom:PRSA:2008}.
By computing the matrix $\mathbf{Q}$ in a sparse fashion, the memory needed can be reduced to approximately $N(2^N-2)$, because from every state there are at most $|V| = N$ transitions possible.
This way, the matrix for a population with 20 individuals needs approximately $168$ MB of memory.
Nevertheless, also the sparse implementation of the matrix grows linear exponentially in population size (with $N 2^N$), and the computational limitations are quickly being reached.

To reduce the working memory needed, one could think of saving the matrix in a file instead of working memory.
Alternatively, one could compute the entries dynamically when needed by the solving of the system of linear equations.
Both approaches, however, would slow down the computation time significantly.
But when working memory is a limiting factor, this would be a feasible possibility to circumvent this limitation.  

We have used the sparse BiCGSTAB method implemented in the eigen3 package \cite{vandervorst:SIAMSSC:1992} for solving the system of linear equations.
This is an iterative method, which means that 
it is prone to rounding errors and the internal error tolerance of the algorithm (usually machine accuracy).
An alternative would be using direct methods. 
However, these solvers need much more memory as well as CPU time for large matrices.

\begin{figure}[h!]
	\centering
    \includegraphics[width=1.0\textwidth]{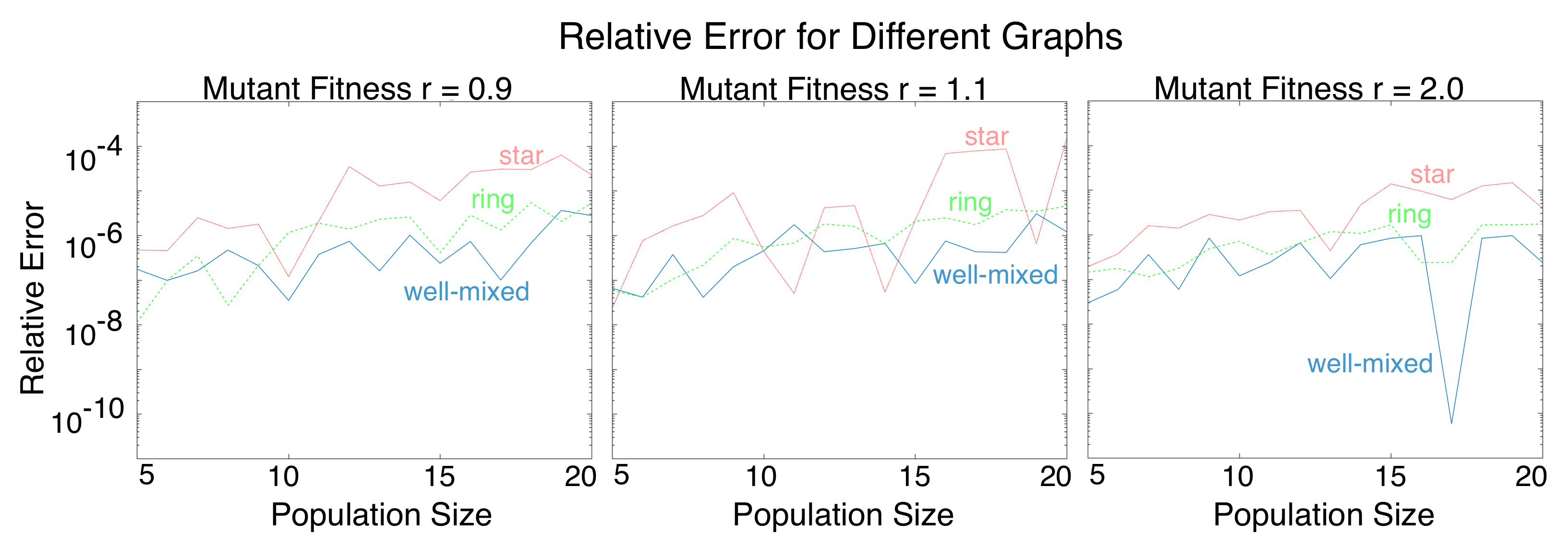}
    \caption{The relative error as the difference of the exact fixation probability and our numerical solution divided by the exact probability for different fitness values of the mutant type.}
   \label{fig:error}
\end{figure}

Fig.~\ref{fig:error} shows the relative error for the fixation probability on three graphs. We see similar errors for different fitness values. The complete graph and the ring show very low relative error of below $10^{-5}$ up to a population size of $N=20$. Even the star only has a relative error of maximum $10^{-4}$. In order to achieve such a precision by doing simulations, one would need an extremely high number of realizations.

\subsubsection{Computational performance}

To test the computational performance, we have computed the fixation probability of random graphs of different sizes for neutral mutants. 

\begin{figure}[h!]
	\centering
    \includegraphics[width=0.8\textwidth]{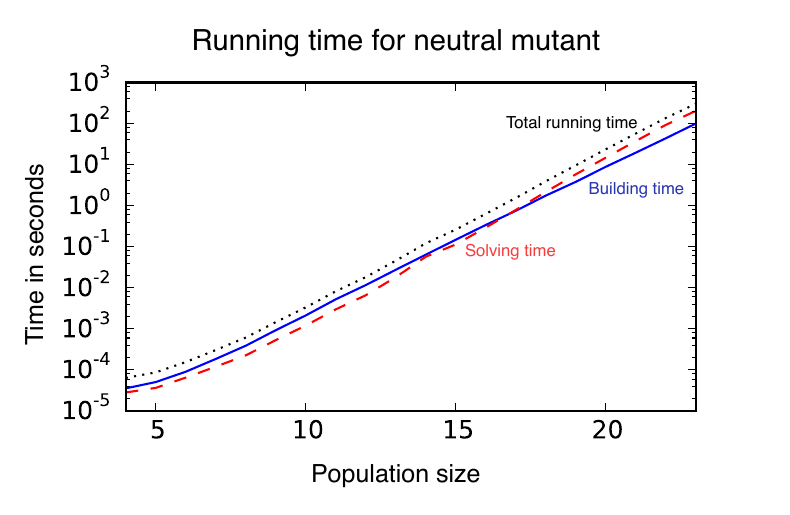}
    \caption{The average times for building the matrix (blue, solid) and solving the system of linear equations (red, dashed) is given here. The sum of the building time and solving time as an indicator of the total running time (black, dotted) is also shown. For each population size between $\{4,\ldots,23\}$ we have created 20 connected, undirected $G(n,p)$ Erd\H{o}s-R\'enyi random graphs with probability $p=0.5$ for each edge to be present. The transition matrix was created for a neutral mutant with fitness $r=1$ and we have solved the system to calculate the fixation probability.}
   \label{fig:performance}
\end{figure}

In Fig.~\ref{fig:performance} the running time in seconds of building the matrix and solving the system of linear equations for all population sizes between 4 and 23 are given.

We have used a MacBook Pro with a 3 GHz Intel Core i7 CPU and 16 GB DDR3 working memory for the tests.
The exact software can be found online \cite{hindersin:URL:2016}. Fig.~\ref{fig:performance} shows that both building time and solving time grow  exponentially in the population size.
Despite the fast growth, for size $N=23$ it takes only of the order of a few minutes to calculate the exact fixation probability.
We also tested different fitness values and graph connectivities which led to similar running times.
Solving the system to compute the unconditional fixation time takes equally long.

\section{Discussion}

In this note we have presented a numerical approach to study evolutionary dynamics of a Moran process on graphs.
Spatial structure can greatly influence both the fixation probability and the fixation time of a mutant population compared to a well-mixed population.
However, dealing with a Moran process on graphs seems to be a difficult undertaking.
Analytical solutions for the fixation probability and time have only been found for a small class of graphs~\cite{lieberman:Nature:2005,broom:PRSA:2008,broom:PRSA:2010,frean:PRSB:2013,askari:PRE:2015}.
Additionally, for general graphs there are approximations~\cite{antal:PRL:2006,alcaldecuesta:BioSys:2015} and a recently developed computational method for the case without selection~\cite{shakarian:BioSys:2013}.
A method that allows one to directly analyze different parameters, such as population size, connectivity, and fitness of the mutants for general graphs may hence be valuable.

Fig.~\ref{fig:performance} shows that our algorithm takes only a few minutes to calculate the exact fixation probability. 
This is remarkable compared to simulating a Moran process on a graph~\cite{santos:PNAS:2006,broom:PRSA:2008}.

This approach makes analyzing evolution on small and medium networks much more interactive, since long-lasting simulations are not necessary anymore.

\subsection*{Acknowledgements}
We thank Kamran Kaveh, Jacob G. Scott  and Jordi Arranz for fruitful discussions and two anonymous reviewers for constructive comments.
Funding from the Max Planck Society is gratefully acknowledged.

\subsection*{Author Contributions}
L.H. and B.B. developed the algorithm. L.H. and M.M. coded the software. All authors wrote the manuscript.

\end{document}